\documentclass[aps,prd,superscriptaddress,floatfix,showpacs]{revtex4}

\usepackage{verbatim}

\usepackage{amsmath}
\usepackage{amsfonts}
\usepackage{amssymb}
\usepackage{graphicx}
\usepackage{slashed}
\usepackage{bm}
\usepackage{xcolor}
\usepackage{epsf}

\usepackage{psfrag}
\usepackage{appendix}

\def\bea{\begin{eqnarray}}
\def\eea{\end{eqnarray}}
\def\beas{\begin{eqnarray*}}
\def\eeas{\end{eqnarray*}}
\def\beqas{\begin{eqnarray*}}
\def\eqas{\end{eqnarray*}}
\def\beq{\begin{equation}}
\def\eeq{\end{equation}}
\def\beqd{\begin{displaymath}}
\def\eeqd{\end{displaymath}}
\def\eqd{\end{displaymath}}

\def\slashchar#1{\setbox0=\hbox{$#1$}
   \dimen0=\wd0
   \setbox1=\hbox{/} \dimen1=\wd1
   \ifdim\dimen0>\dimen1
      \rlap{\hbox to \dimen0{\hfil/\hfil}}
      #1
   \else\begin{eqnarray}
      \rlap{\hbox to \dimen1{\hfil$#1$\hfil}}
      /
   \fi}

\begin{document}

\newcounter{comment}


\newcommand{\CommInlineOG}[1]{{%
		\refstepcounter{comment}%
		\ttfamily\small\textcolor{blue}{\small[$\blacksquare$ \textbf{\underline{Comment}
$\sharp$\thecomment}~by~OG: #1]}}}

\newcommand{\CommInlineBP}[1]{{%
		\refstepcounter{comment}%
		\ttfamily\small\textcolor{red}{\small[$\blacksquare$ \textbf{\underline{Comment}
$\sharp$\thecomment}~by~BP: #1]}}}

\newcommand{\CommInlinePS}[1]{{%
		\refstepcounter{comment}%
		\ttfamily\small\textcolor{orange}{\small[$\blacksquare$ \textbf{\underline{Comment}
$\sharp$\thecomment}~by~PS: #1]}}}

\newcommand{\CommInlineLS}[1]{{%
		\refstepcounter{comment}%
		\ttfamily\small\textcolor{violet}{\small[$\blacksquare$ \textbf{\underline{Comment}
$\sharp$\thecomment}~by~LS: #1]}}}

\newcommand{\CommInlineJW}[1]{{%
		\refstepcounter{comment}%
		\ttfamily\small\textcolor{magenta}{\small[$\blacksquare$ \textbf{\underline{Comment}
$\sharp$\thecomment}~by~JW: #1]}}}


\newenvironment{CommBlockOG}[1][\large$\square$]%
{\refstepcounter{comment}%
\begin{quote}\renewcommand{\baselinestretch}{1}
\color{blue}{#1~ \normalsize \textbf{\underline{Comment} $\sharp$\thecomment~by~OG:~}}}%
{\end{quote}}

\newenvironment{CommBlockBP}[1][\large$\square$]%
{\refstepcounter{comment}%
\begin{quote}\renewcommand{\baselinestretch}{1}
\color{red}{#1~ \normalsize \textbf{\underline{Comment} $\sharp$\thecomment~by~BP:~}}}%
{\end{quote}}

\newenvironment{CommBlockPS}[1][\large$\square$]%
{\refstepcounter{comment}%
\begin{quote}\renewcommand{\baselinestretch}{1}
\color{orange}{#1~ \normalsize \textbf{\underline{Comment} $\sharp$\thecomment~by~PS:~}}}%
{\end{quote}}

\newenvironment{CommBlockLS}[1][\large$\square$]%
{\refstepcounter{comment}%
\begin{quote}\renewcommand{\baselinestretch}{1}
\color{violet}{#1~ \normalsize \textbf{\underline{Comment} $\sharp$\thecomment~by~LS:~}}}%
{\end{quote}}

\newenvironment{CommBlockJW}[1][\large$\square$]%
{\refstepcounter{comment}%
\begin{quote}\renewcommand{\baselinestretch}{1}
\color{magenta}{#1~ \normalsize \textbf{\underline{Comment} $\sharp$\thecomment~by~JW:~}}}%
{\end{quote}}

\title
{Phenomenology of diphoton photoproduction at next-to-leading order}
\author{O.~Grocholski}
\affiliation{Deutsches Elektronen-Synchrotron DESY, Notkestr.
85, 22607 Hamburg, Germany}
\affiliation{National Centre for Nuclear Research (NCBJ), 02-093 Warsaw, Poland}
\author{ B.~Pire }
\affiliation{Centre de Physique Th\'eorique, CNRS, \'Ecole Polytechnique, I.P. Paris, 91128 Palaiseau, France  }
\author{ P.~Sznajder}
\affiliation{National Centre for Nuclear Research (NCBJ), 02-093 Warsaw, Poland}
\author{ L.~Szymanowski}
\affiliation{National Centre for Nuclear Research (NCBJ), 02-093 Warsaw, Poland}
\author{ J.~Wagner}
\affiliation{National Centre for Nuclear Research (NCBJ), 02-093 Warsaw, Poland}

\date{\today}

\begin{abstract}
We develop the analysis of  diphoton exclusive photoproduction in the kinematics where a collinear QCD factorization framework applies, namely nearly forward large invariant mass diphoton production. We work at the leading twist level and at the next-to-leading order (NLO) in the strong coupling constant $\alpha_S$. We compare our predictions for cross-sections with Born order calculations for the experimental conditions accessible to JLab experiments and show the interesting sensitivity of our results to various models of generalized parton distributions (GPDs). The NLO corrections are rather large and negative but do not prevent the studied reaction from being a promising tool for the extraction of $C-$odd GPDs, which do not contribute to either spacelike or timelike deeply virtual Compton scattering amplitudes.
\end{abstract}
\maketitle

\section{Introduction}
\label{sec:intro}

The quest for nucleon tomography in terms of generalized parton distributions (GPDs) \cite{Mueller:1998fv, Ji:1996ek, Ji:1996nm, Radyushkin:1996ru, Radyushkin:1997ki} necessitates the study of as many exclusive processes as possible. Besides the most studied reactions -- deeply virtual Compton scattering (DVCS), timelike Compton scattering (TCS), deeply virtual meson production (DVMP) -- it is now obvious that the special features of diphoton production,
\begin{equation}
\gamma(q,\epsilon) + N(p_1,s_1) \rightarrow \gamma(q_1,\epsilon_1) +  \gamma(q_2,\epsilon_2)+ N(p_2,s_2)\,,
\label{process}
\end{equation}
makes it a necessary tool for the hadronic program. Namely, this reaction allows us to access the charge-conjugation odd quark (often called valence) GPDs, which are decoupled from DVCS, TCS and DVMP of neutral vector mesons~\footnote{Present phenomenological studies of pseudoscalar mesons production, which is theoretically the easiest place to access the charge conjugation odd GPDs, indicate a quite important higher twist component of their amplitudes, delaying a leading twist extraction to be meaningful at present energies. Production of these mesons is also mainly sensitive to the GPDs $\widetilde{H}$ and $\widetilde{E}$, whereas diphoton photoproduction to the GPDs $H$ and $E$. Charged vector meson production is rather difficult to experimentally access. We also note that charge conjugation odd GPDs are probed by elastic form factors. There however, only limited information about these GPDs is accessible, namely, only their zero-th Mellin moments can be measured.}. 

In our previous works~\cite{Pedrak:2017cpp, Pedrak:2020mfm} we studied the photoproduction of a large mass diphoton on a nucleon target at leading order (LO) in the QCD coupling constant $\alpha_s$ in the kinematical domain suitable for the factorization of amplitudes into GPDs and hard scattering parts. Next-to-leading order (NLO) description has been added in our recent work~\cite{Grocholski:2021man}, where we mostly focused on the development and presentation of complex analytic expressions. The comparison between LO and NLO results at the level of amplitude and cross-section values is the subject of the present article. The subject is important, as numerical estimates of NLO effects are still rare in the field of GPD studies. Diphoton photoproduction also offers a unique set of conditions to study higher order effects. Namely, gluons do not contribute here directly to the amplitudes at both LO and NLO (in contrast to DVCS and TCS, which \emph{nota bene} share a very similar structure of the description at both orders~\cite{Mueller:2012sma,Grocholski:2019pqj}), and there are no non-perturbative ingredients other than GPDs (in contrast to DVMP). In addition, the process does not probe so-called  D-terms \cite{Polyakov:1999gs}, making its interpretation exceptionally simple in phenomenology studies (in contrast to all other aforementioned exclusive processes).

The article is organised as follows. A concise description of the scattering amplitudes, including all definitions  of kinematic variables, is provided in Sect.~\ref{sec:theory}. Numerical estimates for both the amplitudes and cross-sections for the case of  unpolarised target and beam are given in Sect.~\ref{sec:phenoU}. We show the effects of an initial photon linear polarization in Sect.~\ref{sec:phenoL}, while in Sect.~\ref{sec:phenoT} we study the asymmetry for a transversally polarised target, which has an interesting property to vanish at leading order, but to be sizeable at NLO. A brief summary is given in Sect.~\ref{sec:summary}.  

\begin{figure}[!ht]
\begin{center}
\includegraphics[width=0.30\textwidth]{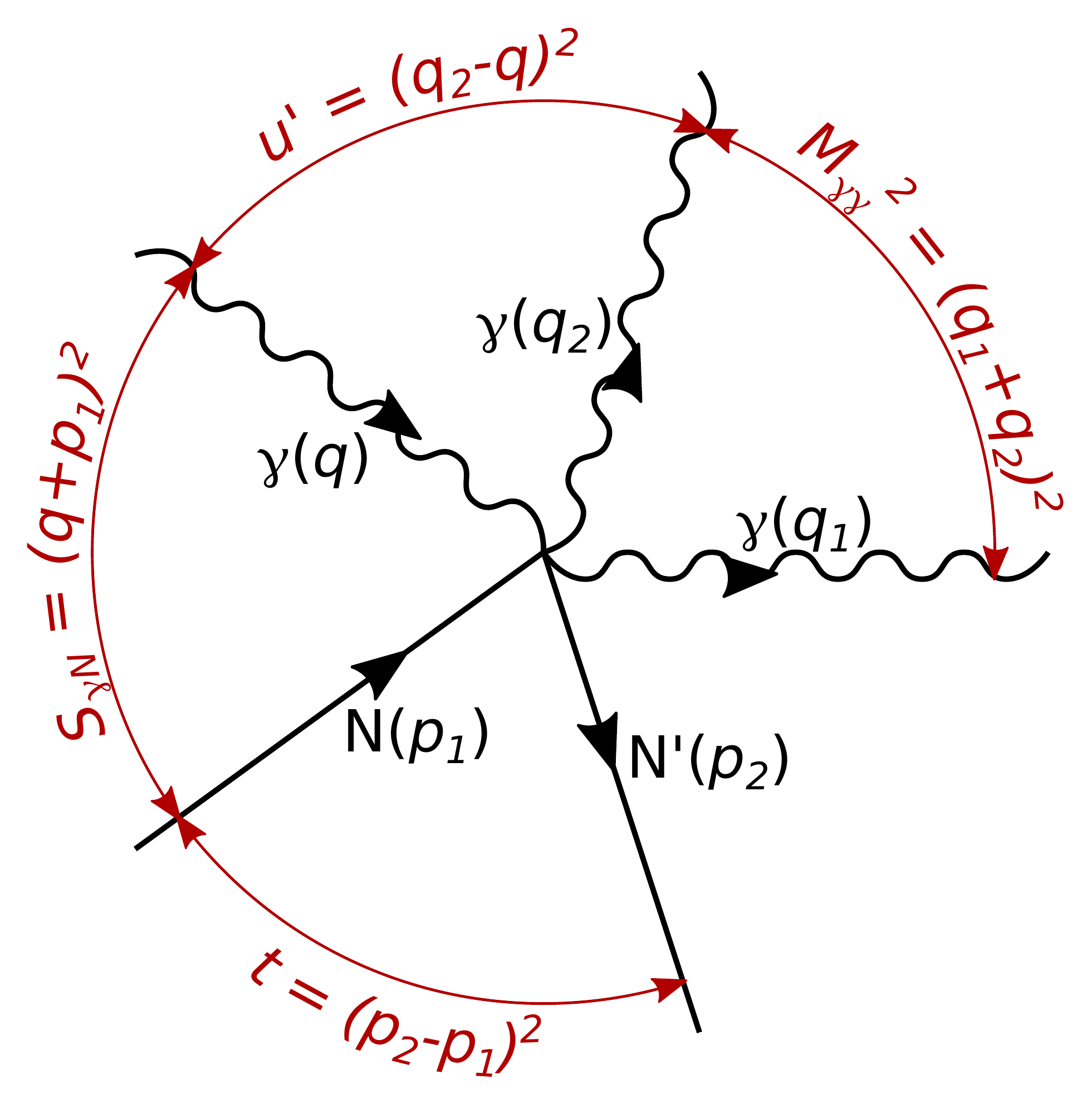}
\caption{Kinematics of diphoton photoproduction.}
\label{fig:intro:process}
\end{center}
\end{figure}

\section{The scattering amplitude}
\label{sec:theory}

Let us briefly recall the kinematics and the formulae for the scattering amplitude of the process (\ref{process}), that we derived in our previous works \cite{Pedrak:2017cpp,  Grocholski:2021man}.
We parametrize 
the momenta in the 
Sudakov basis using two lightlike vectors $p$ and $n$, defined such as $p^2 = n^2 = 0$ and $s= 2 p\cdot n$, which gives:
\begin{equation}
\begin{aligned}
&p_1^\mu = (1+\xi)p^\mu + \frac{M^2}{s(1+\xi)}n^\mu, \qquad p_2^\mu = (1-\xi)p^\mu + \frac{M^2 + \vec{\Delta}_\perp^2}{s(1-\xi)}n^\mu + \Delta_\perp^\mu, \qquad q^\mu = n^\mu,
\\
&q_1^\mu = \alpha n^\mu + \frac{(\vec{p}_\perp-\frac{1}{2}\vec{\Delta}_\perp)^2}{\alpha s} p^\mu + p_\perp^\mu - \frac{1}{2}\Delta_\perp^\mu,
\\
&q_2^\mu = \bar{\alpha} n^\mu + \frac{(\vec{p}_\perp+\frac{1}{2}\vec{\Delta}_\perp)^2}{\bar{\alpha} s} p^\mu - p_\perp^\mu - \frac{1}{2}\Delta_\perp^\mu.
\end{aligned}
\end{equation}
Here, $M$ is the mass of the nucleon, $\xi$ is the skewness variable and $\vec{v}\cdot \vec{u}$ denotes the Euclidean product of transverse vectors  ($\vec{v}^{\phantom{.}2} \geq 0$). The simplified kinematical relations used in the hard subprocess amplitude  read:
\begin{equation}
\begin{aligned} \label{eq-kin-simplified}
 \tau = \frac{M_{\gamma\gamma}^2 - t}{S_{\gamma N} - M^2}&, \qquad \xi = \frac{\tau}{2-\tau}, \\
   M_{\gamma\gamma}^2 = \frac{\vec{p}_\perp^{\:2}}{\alpha \bar{\alpha}}, \qquad  -t' = \frac{\vec{p}_\perp^{\:2}}{\alpha}&, \qquad -u' = \frac{\vec{p}_\perp^{\:2}}{\bar{\alpha}}, \qquad \alpha + \bar{\alpha} = 1.
\end{aligned}
\end{equation}
We write the amplitude in terms of the invariants, see also Fig. \ref{fig:intro:process}:
\begin{equation}
S_{\gamma N} = (q + p_1)^2, \quad t = (p_2 - p_1)^2, \quad M_{\gamma\gamma}^2 = (q_1 + q_2)^2, \quad u' = (q_2 - q)^2.
\end{equation}
Polarization vectors of the photons are written in the $\epsilon \cdot p =0$ gauge 
as:
\begin{equation}
\begin{aligned}
&\epsilon^\mu(q_1) = \epsilon_\perp^\mu(q_1) - \frac{2\epsilon_\perp\big{(}p_\perp - \frac{1}{2}\Delta_\perp\big{)}}{\alpha s}p^\mu,
\\
&\epsilon^\mu(q_2) = \epsilon_\perp^\mu(q_2) + \frac{2\epsilon_\perp\big{(}p_\perp + \frac{1}{2}\Delta_\perp\big{)}}{\bar{\alpha} s}p^\mu,
\\
&\epsilon^\mu(q) = \epsilon_\perp^\mu(q).
\end{aligned}
\label{eq-polarizations}
\end{equation}
The unpolarised cross-section reads:
\begin{equation}
\frac{d^{3}\sigma}{\mathrm{d}M^2_{\gamma \gamma}\,\mathrm{d}t\,\mathrm{d}(-u')} = \frac{1}{2}\,\frac{1}{(2\pi)^3\, 32\,S^2_{\gamma\,N}M^2_{\gamma\,\gamma}}\,\sum_{\lambda,\lambda_1,\lambda_2,s_1,s_2}\,\frac{|{\cal T}|^2}{4}\,,
\end{equation}
where the sum runs over the polarisation states of the initial photon and nucleon ($\lambda$, $s_{1}$), and the outgoing photons and nucleon ($\lambda_{1}$, $\lambda_{2}$, $s_{2}$).

The factorization property \cite{Grocholski:2021man} of the scattering amplitude  $\mathcal{T} $ allows us to write it as:
\begin{eqnarray}
&\mathcal{T} & =
\frac{1}{2}\;\frac{1}{2P^+}
\int_{-1}^1 dx\sum_q\left[
T^q(x,\xi)\left(
 H^q(x,\xi)\bar{U}(p_2, s_2)\not{n} U(p_1,s_1)+
 E^q(x,\xi)\bar{U}(p_2, s_2)\frac{i\sigma^{\mu\nu}\Delta_\nu n_\mu}{2M}U(p_1, s_1)
\right)+
\right.\nonumber \\
&&\left. ~~~~~~~~~~~~~~~~~~~~~
+\widetilde T^q(x,\xi)\left(
\widetilde{H}^q(x,\xi)\bar{U}(p_2, s_2)\not{n}\gamma^5 U(p_1, s_1)+
\widetilde{E}^q(x,\xi)\bar{U}(p_2, s_2)\frac{i\gamma_5(\Delta\cdot n)}{2M}U(p_1, s_1)
\right)
\right]\, .
\label{eq7}
\end{eqnarray}
Eq. (\ref{eq7}) contains the axial part proportional to the $\widetilde H$ and $\widetilde  E$ GPDs, which however gives a quite negligible contribution to the cross-section (already at the  LO level) and we shall forget it in the following.

The coefficient functions are: 
\begin{eqnarray}
T^q &=&
C_0^q 
+ C_1^{q}
+\log
\left(
\frac{M^2_{\gamma\gamma}}{\mu_F^2}
\right) C_{coll}^{q} 
\,,
\label{eq:coeff}
\end{eqnarray}
with the LO part \cite{Pedrak:2017cpp}: 
\begin{equation}\label{eq-Cq0-recap}
\begin{aligned}
	&C^q_0 (x, \xi) = \frac{ie_q^3 A }{ M^2_{\gamma\gamma} \alpha \bar{\alpha} } \: \mathrm{Im} \Big( \frac{1}{x+\xi - i0}  \Big) + ( x \rightarrow -x)\,,
	\end{aligned}
\end{equation}
where:
\begin{equation}
\begin{aligned}	    
A =	\bigg[(\alpha - \bar{\alpha} )\big( \vec{\epsilon^*}_t(\mathbf{q}_1) \vec{\epsilon^*}_t(\mathbf{q}_2) \big) \big( \vec{p}_t \vec{\epsilon}_t (\mathbf{q}) \big)  - \big(  \vec{p}_t \vec{\epsilon^*}_t(\mathbf{q}_1) \big) \big(  \vec{\epsilon}_t (\mathbf{q}) \vec{\epsilon^*}_t(\mathbf{q}_2) \big) + \big(  \vec{p}_t \vec{\epsilon^*}_t(\mathbf{q}_2) \big) \big(  \vec{\epsilon}_t(\mathbf{q}) \vec{\epsilon^*}_t(\mathbf{q}_1) \big) \bigg]  .
\end{aligned}
\end{equation}
and the collinear evolution part
\begin{eqnarray}
C_{coll}^{q}(x') &=& C_0^q(x) \otimes K_{NS}^{qq}(x,x')~~,
\label{eq:div_cancel}
\end{eqnarray}
where $K_{NS}^{qq}(x,x')$ is the non-singlet GPD evolution kernel~\cite{Diehl:2003ny}. $C_1^{q}$ is the NLO scale-invariant part, which full form is given in \cite{Grocholski:2021man}.

The Compton form factors (CFFs) \footnote{
In this work we use the name ``Compton form factors'' for functions depending on all invariants and polarizations of photons.},  which we consider below, are thus defined as:
\begin{equation}
    {\cal H}^q(\xi) = \int_{-1}^1 dx \;T^q(x,\xi)\,H^q(x,\xi)\;,\;\;\;\; {\cal E}^q(\xi) = \int_{-1}^1 dx\; T^q(x,\xi)\, E^q(x,\xi)\;,
\end{equation}
and 
\begin{equation}
{\cal H}(\xi) = \displaystyle\sum_q {\cal H}^q(\xi),  \;\;\;{\cal E}(\xi) = \displaystyle\sum_q {\cal E}^q(\xi)\,.
\end{equation}

\section{Compton form factors and cross-sections for unpolarized target}

Let us now present our estimates for Compton form factors and cross-sections, and show the effects of including NLO corrections to the Born order estimates. The code for the numerical evaluation of both CFFs and cross-sections has been implemented in the open source PARTONS framework \cite{Berthou:2015oaw}. We note that because of the complexity of NLO expressions, CFF convolutions are evaluated numerically, including an explicit treatment of the Feynman epsilon prescription and multi-dimensional integration performed with Monte Carlo methods. The later is necessary because of the cumbersome integrals emerging at NLO (see $\mathcal{F}_{nab}(x,\xi,\{\beta_{i}\})$ and $\mathcal{G}(x,\xi,\{\beta_{i}\})$ defined in Eqs.~(41)~and~(42), respectively, of Ref.~\cite{Grocholski:2021man}).
As a consequence, our numerical estimates for NLO are very costly in terms of computing time, and sometimes exhibit a visible numerical noise. This prevents us for instance from integrating differential cross-sections, with the exception of $\phi$ angle, for which the integration is straightforward. We therefore only show results for either four-fold, $\mathrm{d}^4\sigma/(\mathrm{d}t\,\mathrm{d}u'\,\mathrm{d}M_{\gamma\gamma}^2\,\mathrm{d}\phi)$, or three-fold, $\mathrm{d}^3\sigma/(\mathrm{d}t\,\mathrm{d}u'\,\mathrm{d}M_{\gamma\gamma}^2)$, differential cross-sections. A minor advantage is that without the integration we can show CFFs corresponding to the presented cross-sections straightforwardly. 

In most cases  we chose the kinematics associated to the domain covered by JLab experiments. We explain this choice in the following section. Since the axial part of the amplitude gives a subdominant contribution to the cross-section at LO, and since we are mostly interested in showing the effect of NLO correction to the dominant part, we neglect both $\widetilde H$ and $\widetilde E$ GPDs in our analysis. The $H$ and $E$ GPDs  enter the evaluation of all observables presented in this article, however for brevity CFF $\mathcal{E}$ related to the $E$ GPD is only shown in the case of transversally polarised target, where it contributes significantly. In the following predictions we keep the factorization and renormalization scales equal to the invariant mass of the photon pair $\mu_R=\mu_F=M_{\gamma\gamma}$.

\begin{figure}[!ht]
\begin{center}
\includegraphics[width=0.3\textwidth]{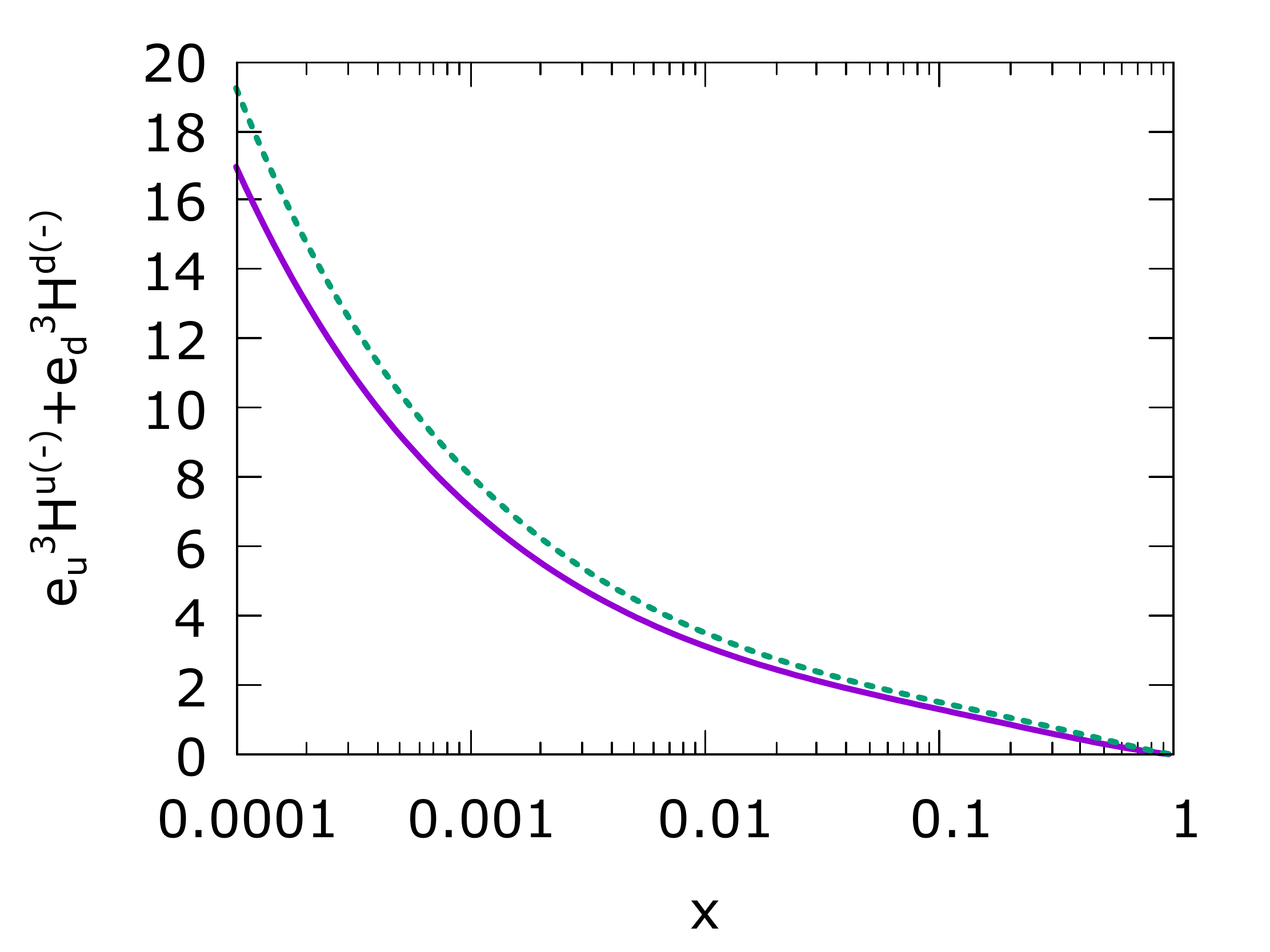}
\includegraphics[width=0.3\textwidth]{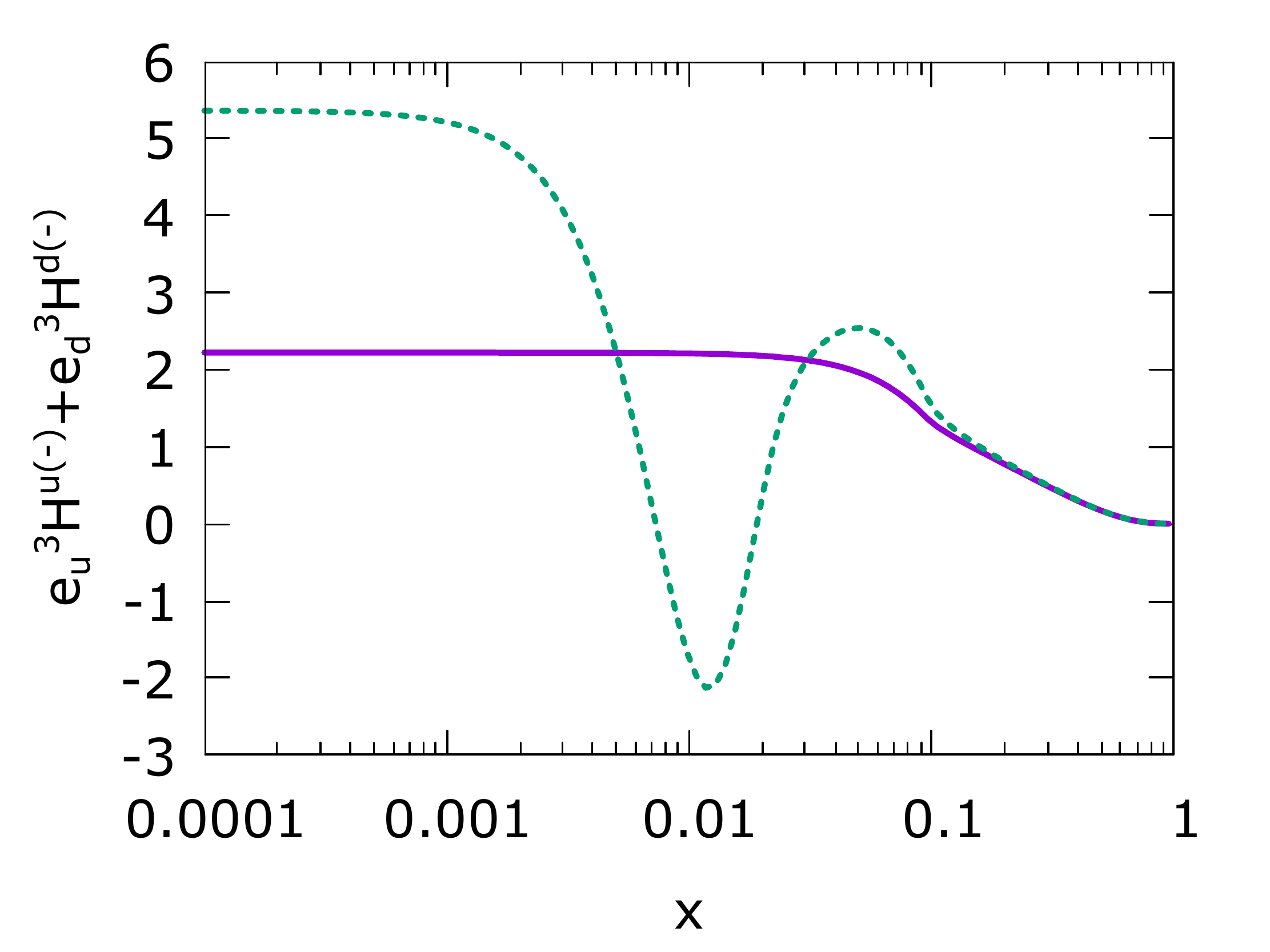}
\includegraphics[width=0.3\textwidth]{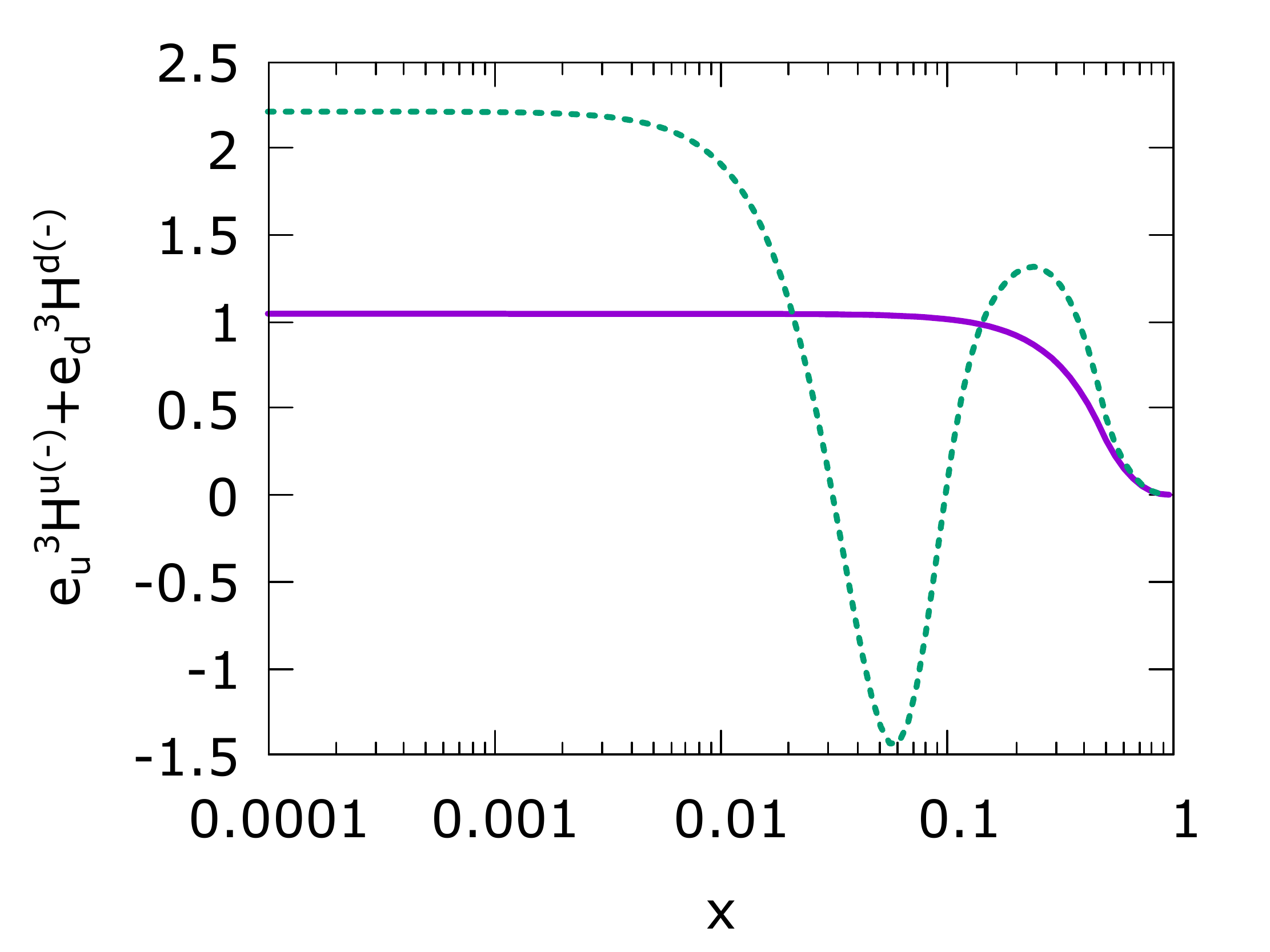}
\includegraphics[width=0.3\textwidth]{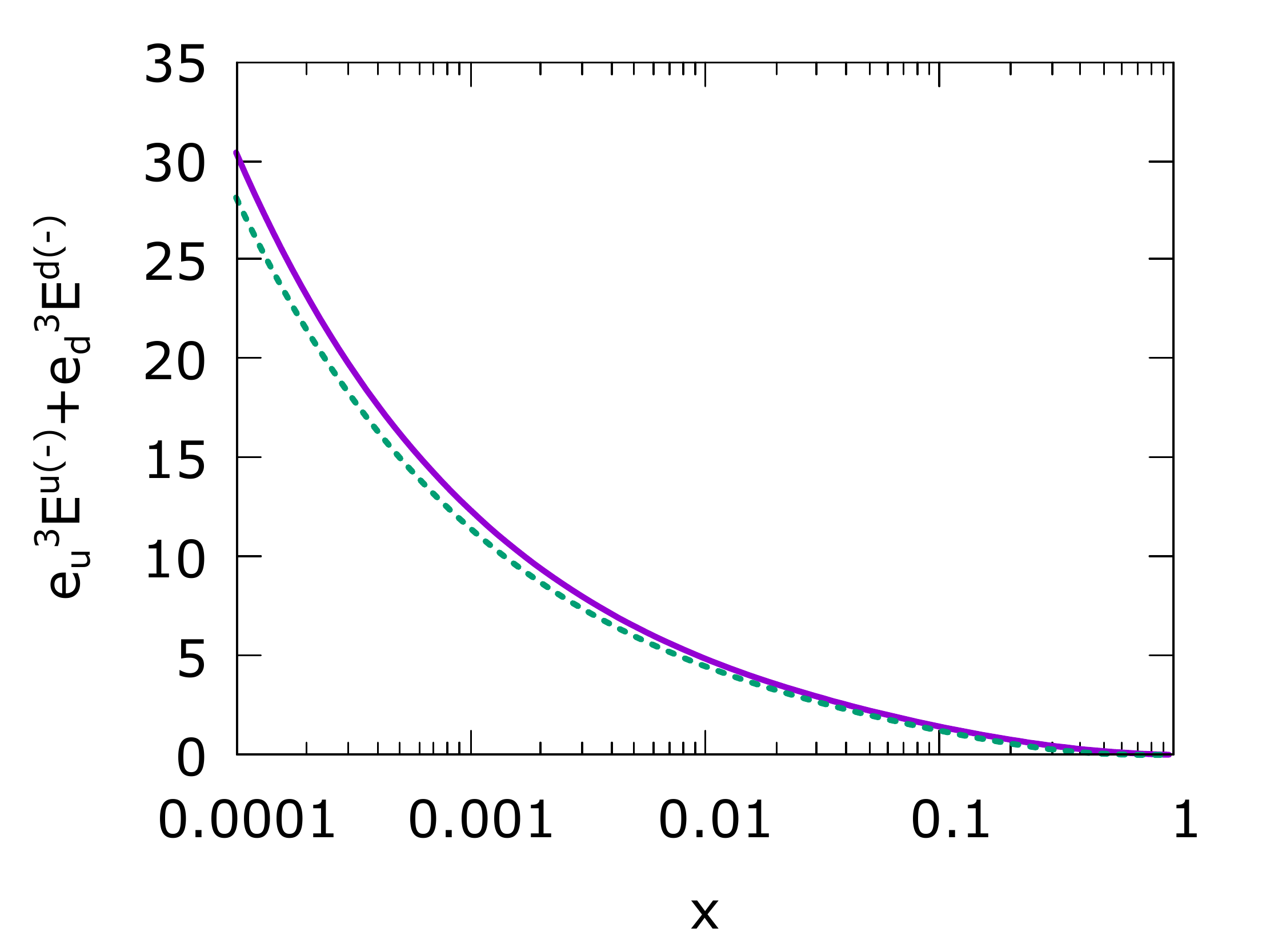}
\includegraphics[width=0.3\textwidth]{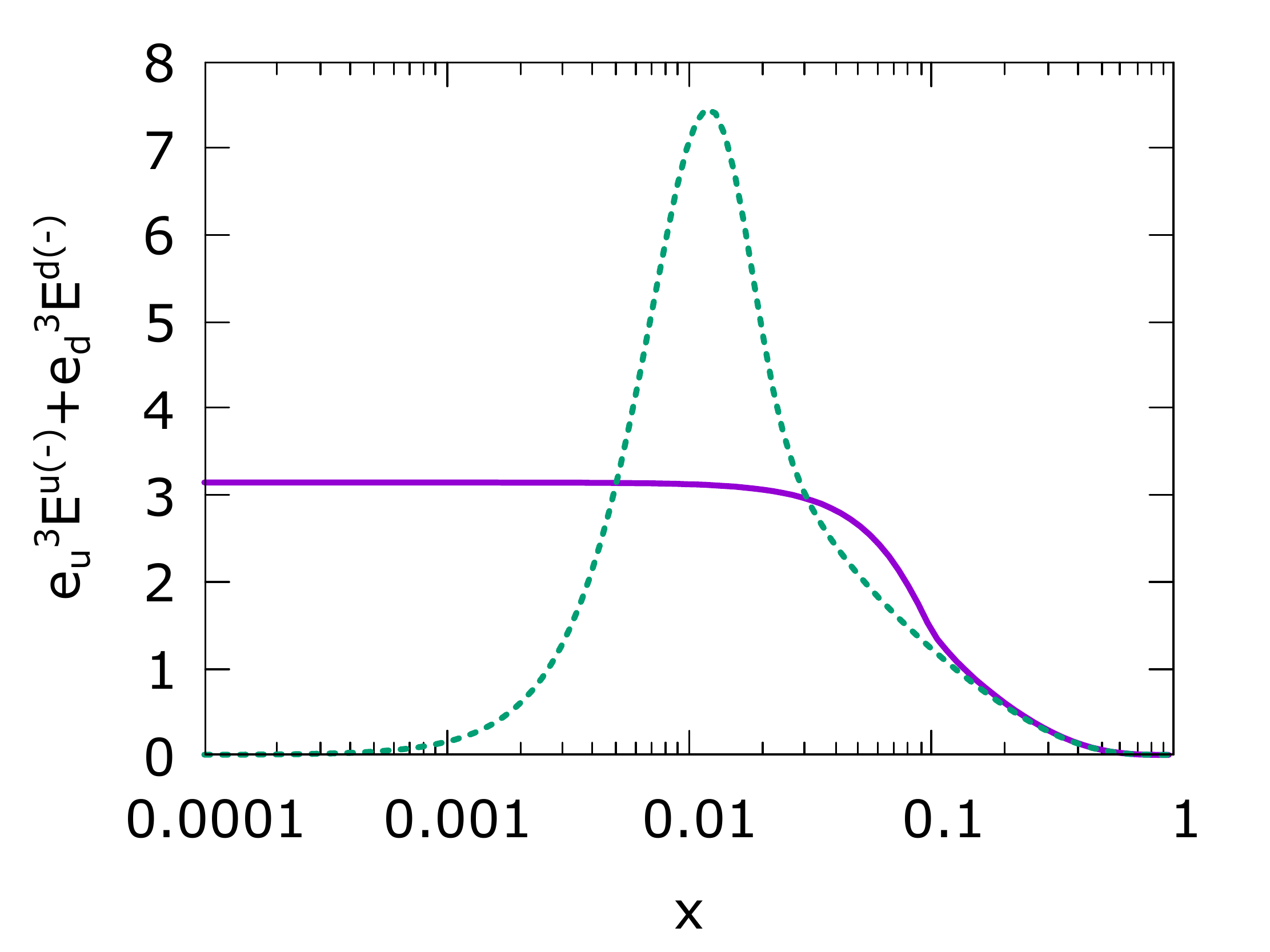}
\includegraphics[width=0.3\textwidth]{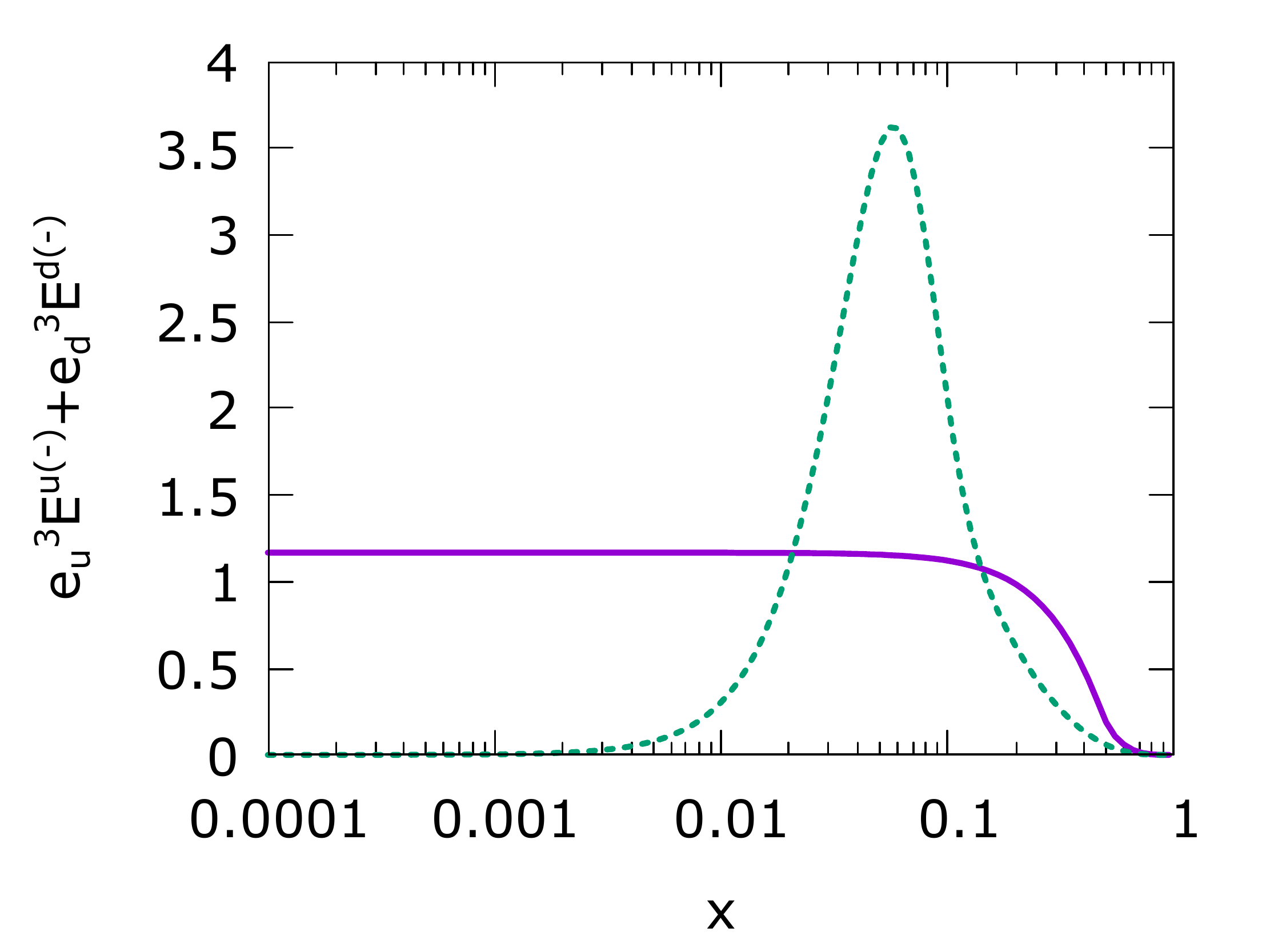}
\caption{Distributions of (top row) $e_u^3 H^{u\,(-)}(x,\xi) + e_d^3 H^{d\,(-)}(x,\xi)$ and (bottom row) $e_u^3 E^{u\,(-)}(x,\xi) + e_d^3 E^{d\,(-)}(x,\xi)$ for (left column) $\xi=x$, (middle column) $\xi=0.1$ and (right column) $\xi=0.5$. The solid magenta and dotted green curves are for GK and MMS GPD models, respectively. The C-odd parts of GPDs are defined as: $H^{q\,(-)}(x,\xi) = H^{q}(x,\xi) + H^{q}(-x,\xi)$ and $E^{q\,(-)}(x,\xi) = E^{q}(x,\xi) + E^{q}(-x,\xi)$ \cite{Diehl:2003ny}. The plots are for $t = -0.1~\mathrm{GeV}^2$ and the scale $\mu_F^2 = 4~\mathrm{GeV}^2$.}
\label{fig:gpd}
\end{center}
\end{figure}
 
Our results use the GK \cite{Goloskokov:2007nt, Kroll:2012sm} and MMS \cite{Mezrag:2013mya} GPD models illustrated in Fig. \ref{fig:gpd}. These models were chosen because of differences in their construction~--~GK is based on two-component, while MMS on one-component double distribution~\cite{Belitsky:2005qn}~--~allowing us to show the sensitivity of results to a specific choice. The models are also constrained by different sets of data covering different kinematical domains: GK by DVMP of vector mesons at moderately low-$\xi$ values, while MMS by DVCS in high-$\xi$ domain. To relate DVMP data with GPDs GK used the so-called modified perturbative approach \cite{Botts:1989kf,Li:1992nu,Jakob:1993iw} with LO coefficient functions.
In the case of MMS the standard LO description of DVCS was used, see Ref. \cite{Kroll:2012sm} and references therein.
Both models roughly reproduce data for exclusive processes sensitive to the C-even part of GPDs, but were not much tested for the C-odd part. 
Because of all aforementioned differences and possible inconsistencies in LO/NLO description, our curves should not be understood as definite predictions for the observables of diphoton photoproduction, but mainly as a proof of the importance of the NLO extraction of GPDs from data for this and other exclusive processes. 

\subsection{Unpolarized cross-sections}
\label{sec:phenoU}
\begin{figure}[!ht]
\begin{center}
\includegraphics[width=0.5\textwidth]{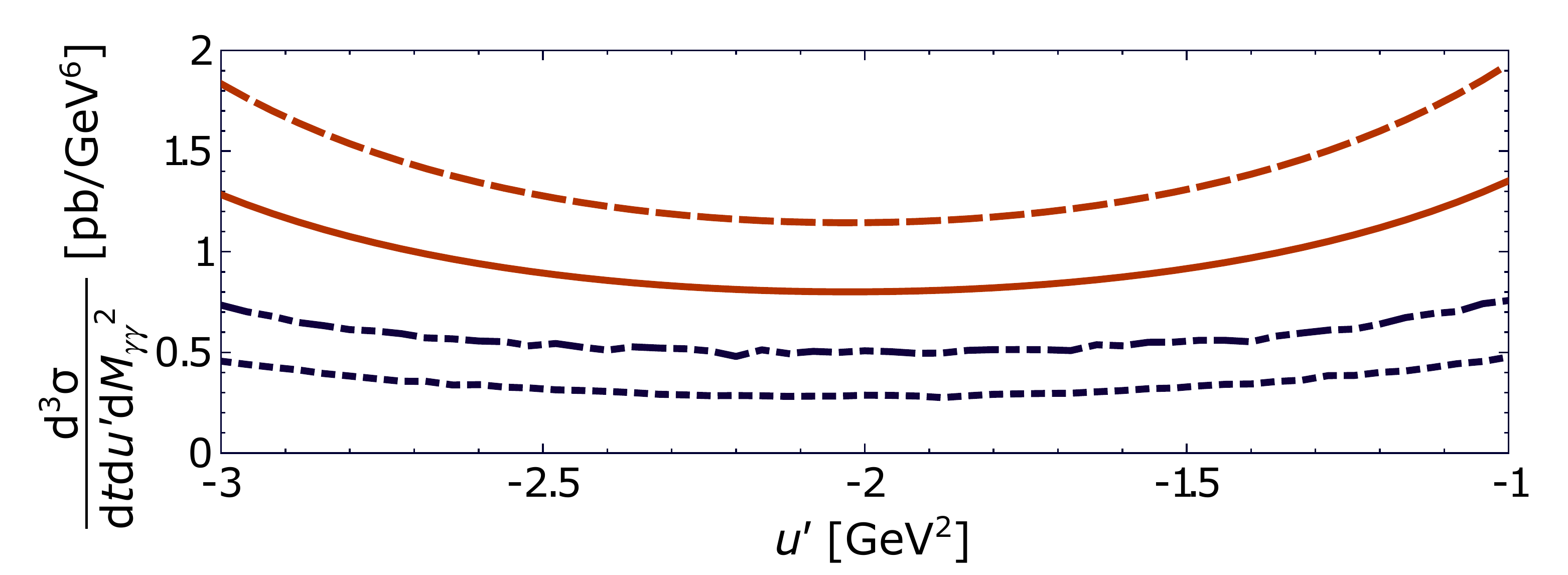}
\includegraphics[width=1.0\textwidth]{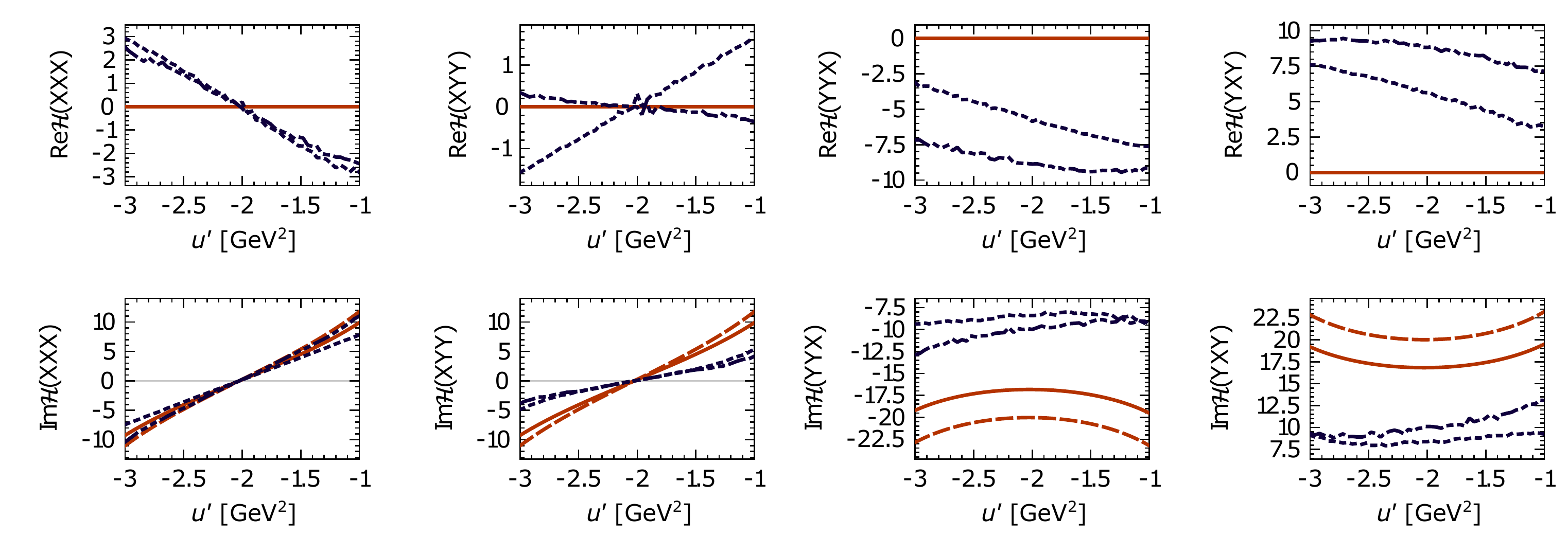}
\caption{(top) Differential cross-section as a function of $u'$. (bottom) Compton form factors $\mathcal{H}$ as a function of $u'$ for various combinations of polarisations. The combinations for which Compton form factors explicitly vanish are not shown. (both) Leading order calculation is denoted by the solid (dashed) red line, while next-to-leading one by dotted (dash-dotted) blue line for GK (MMS) GPD model. The plots are for $S_{\gamma N} = 20~\mathrm{GeV}^2$, $M_{\gamma\gamma}^2 = 4~\mathrm{GeV}^2$ (which corresponds to $\xi \approx 0.12$) and $t=t_{0} \approx -0.05~\mathrm{GeV}^2$.}
\label{fig:plot_1}
\end{center}
\end{figure}

 We start the presentation of our results with Fig. \ref{fig:plot_1}, where the three-fold differential cross-section for unpolarised proton target is shown  as a function of $u'$ at  $S_{\gamma N} = 20~\mathrm{GeV}^2$ and $M_{\gamma\gamma}^2 = 4~\mathrm{GeV}^2$ (which corresponds to $\xi \approx 0.12$) and at $t=t_{0}=-4 \xi^2 M^2/(1-\xi^2) \approx -0.05~\mathrm{GeV}^2$. We plot the curves for both LO and NLO coefficient functions, and for both GK and MMS models. The plot for the cross-section is supplemented by plots showing the real and imaginary parts of CFF $\mathcal{H}$ for various combinations of linear polarization of photons, which are indicated in the vertical labels. We mark these combinations in the following way: $(ABC)$, where $A,B,C \in \{X, Y\}$ denote polarization states of incoming and two outgoing photons, respectively. Here, $X$ means that the transverse part of the polarization vector of a considered photon is in the same direction as momentum $\vec{p}_\perp$, while $Y$ that it is orthogonal, see Eqs. \eqref{eq-kin-simplified} and \eqref{eq-polarizations}. Polarization combinations for which CFFs strictly vanish for both LO and NLO are not presented. From Fig. \ref{fig:plot_1} we may conclude that the inclusion of the NLO contribution results in a significant reduction of the cross-sections. This clearly indicates the importance of including higher-order corrections in order to get a reliable extraction of GPDs. We stress that a similar conclusion has been already drawn also for other exclusive processes, see \emph{e.g.} Ref. \cite{Moutarde:2013qs}. The importance of higher order corrections has been also proven many times in the case of PDF phenomenology. Also note that the real part of CFFs at LO vanish explicitly.
 
\begin{figure}[!ht]
\begin{center}
\includegraphics[width=0.5\textwidth]{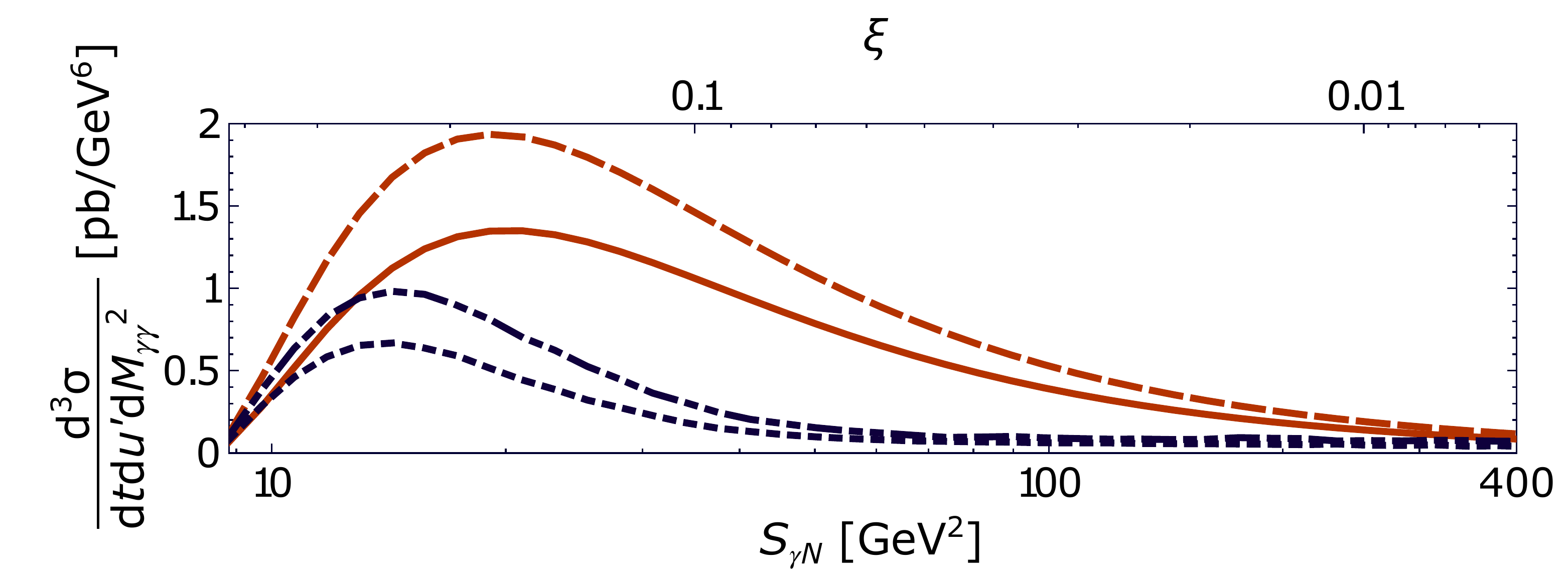}
\includegraphics[width=1.0\textwidth]{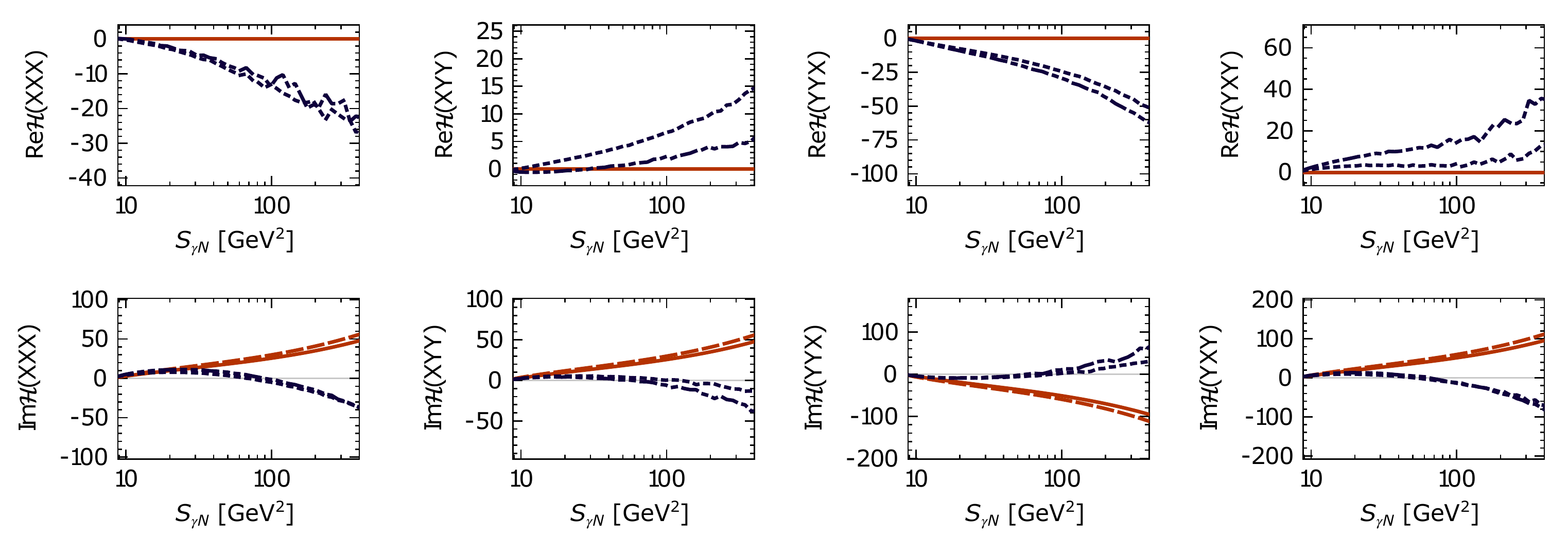}
\caption{(top) Differential cross-section as a function of $S_{\gamma N}$ (bottom axis) and the corresponding $\xi$ (top axis). (bottom) Compton form factors $\mathcal{H}$ as a function of $S_{\gamma N}$ for various combinations of polarisations. The plots are for $M_{\gamma\gamma}^2 = 4~\mathrm{GeV}^2$, $t=t_{0}$ and $u' = -1~\mathrm{GeV}^2$. For further description see the caption of Fig. \ref{fig:plot_1}.}
\label{fig:plot_2_uprim_m1}
\end{center}
\end{figure}

\begin{figure}[!ht]
\begin{center}
\includegraphics[width=0.5\textwidth]{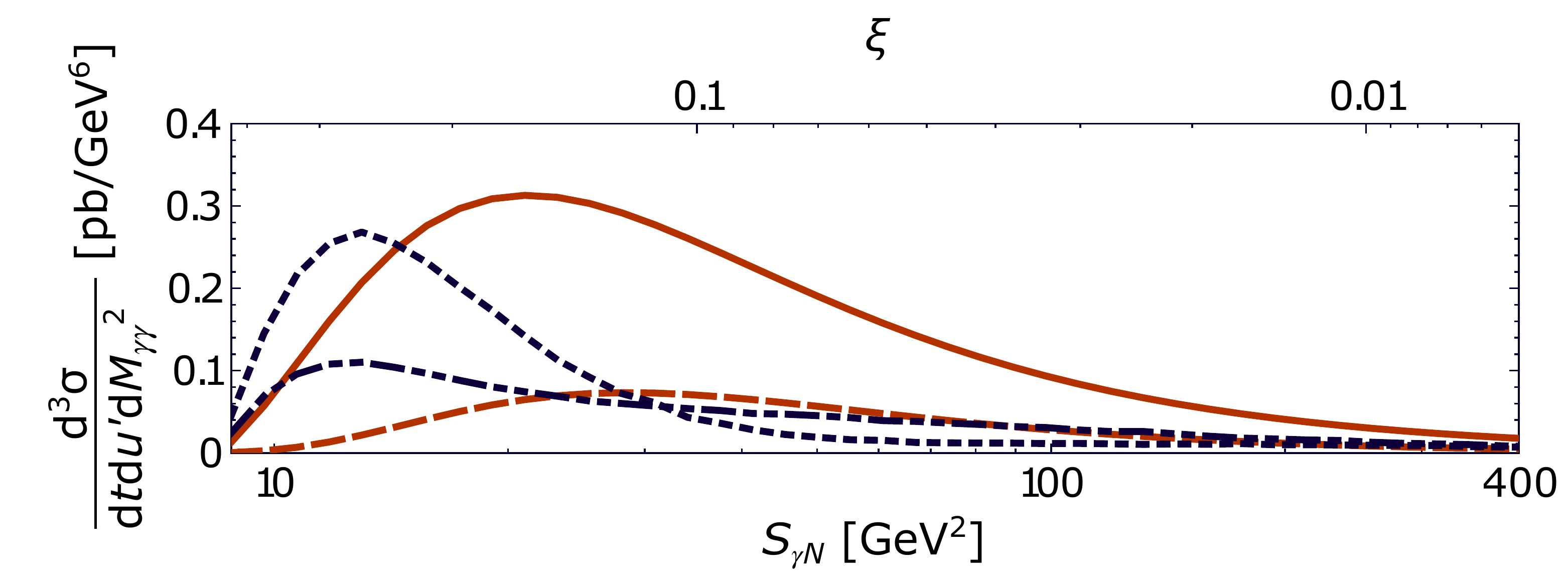}
\includegraphics[width=1.0\textwidth]{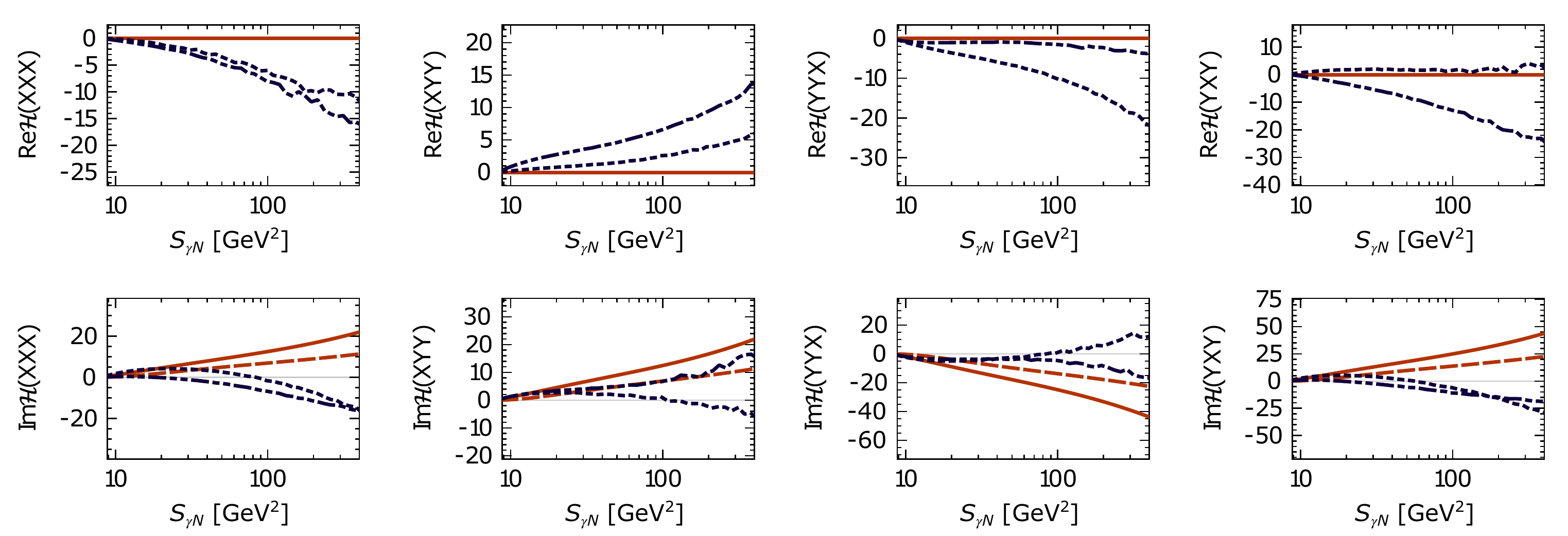}
\caption{The same as Fig. \ref{fig:plot_2_uprim_m1}, but for neutron target.}
\label{fig:plot_2_uprim_m1_neutron}
\end{center}
\end{figure}

Figure \ref{fig:plot_2_uprim_m1} shows the same quantities as Fig. \ref{fig:plot_1}, but this time as a function of $S_{\gamma N}$ at $M_{\gamma\gamma}^2 = 4~\mathrm{GeV}^2$, $t=t_{0}$ and $u' = -1~\mathrm{GeV}^2$. The corresponding values of $\xi$ are indicated in the upper axis. The main conclusion that we can draw here is that according to the used GPD models the most preferable kinematics to measure diphoton photoproduction is the high-$\xi$ region, namely the domain covered by JLab experiments. This comes as no surprise, as the C-even part of GPDs probed by the process can be associated to the valence sector. 

Our results for the neutron target are show in Fig. \ref{fig:plot_2_uprim_m1_neutron}.  To make these curves we used GK and MMS models for proton GPDs and the isospin dictated change of quark contributions: $H^u \rightarrow H^d $ and $H^d \rightarrow  H^u$ (and similar changes for other types of GPDs). The conclusions are similar as in the case of the proton target, however with a less favourable cross-section, being smaller by a factor of the order $5$. The origin of this factor is clearly the small value of the ratio $(8 H^d - H^u)/(8 H^u - H^d)$, which appears in the ratio of Compton form factors.

\begin{figure}[!ht]
\begin{center}
\includegraphics[width=0.5\textwidth]{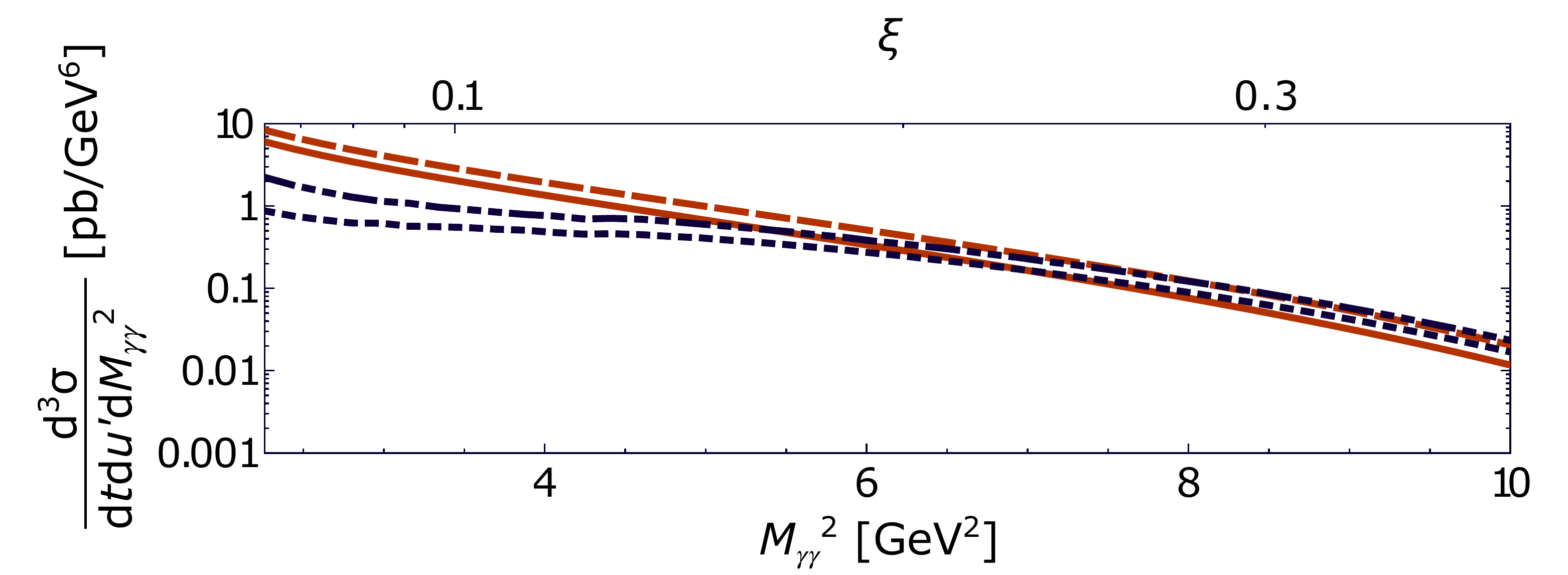}
\includegraphics[width=1.0\textwidth]{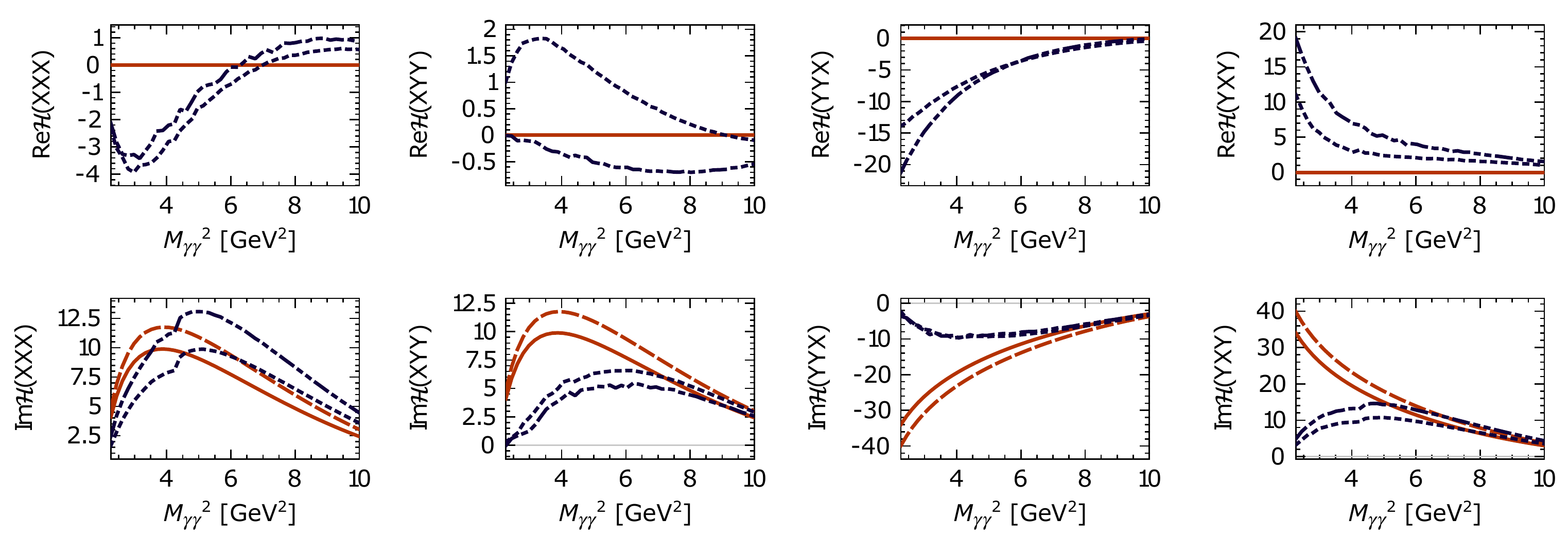}
\caption{(top) Differential cross-section as a function of $M_{\gamma\gamma}^2$ (bottom axis) and the corresponding $\xi$ (top axis). (bottom) Compton form factors $\mathcal{H}$ as a function of $M_{\gamma\gamma}^2$ for various combinations of polarisations. The plots are for $S_{\gamma N} = 20~\mathrm{GeV}^2$, $t=t_{0}$ and $u' = -1~\mathrm{GeV}^2$. For further description see the caption of Fig. \ref{fig:plot_1}.}
\label{fig:plot_3_uprim_m1}
\end{center}
\end{figure}

Returning back to the proton target case, on Fig. \ref{fig:plot_3_uprim_m1}, we show the $M_{\gamma\gamma}^2$-dependence at $S_{\gamma N} = 20~\mathrm{GeV}^2$, $t=t_{0}$ and $u' = -1~\mathrm{GeV}^2$. As expected, the cross-section dies out quickly as $M_{\gamma\gamma}^2$ is increasing. We also see that NLO corrections become less important for high values of $M_{\gamma\gamma}^2$ (\emph{i.e.} high values of the scale). This expected result supports  the correctness of both analytical and numerical aspects of our analysis. Note that our calculations use the so-called forward evolution of GPDs, that is, the scale evolution of GPDs is restricted to the DGLAP evolution of the PDF part entering the double distribution. The effect of this approximation was checked by the comparison with estimates obtained with the full GPD evolution, using the algorithm of Ref. \cite{Vinnikov:2006xw}. The inclusion of this algorithm slowed down the computation of CFFs even further, forcing us to give up much of the numerical precision. Within the numerical noise  we do not notice any significant difference between results obtained with either forward or full GPD evolution.

\subsection{Initial photon linear polarization effects}
\label{sec:phenoL}

\begin{figure}[h]
\begin{center}
\includegraphics[width=0.5\textwidth]{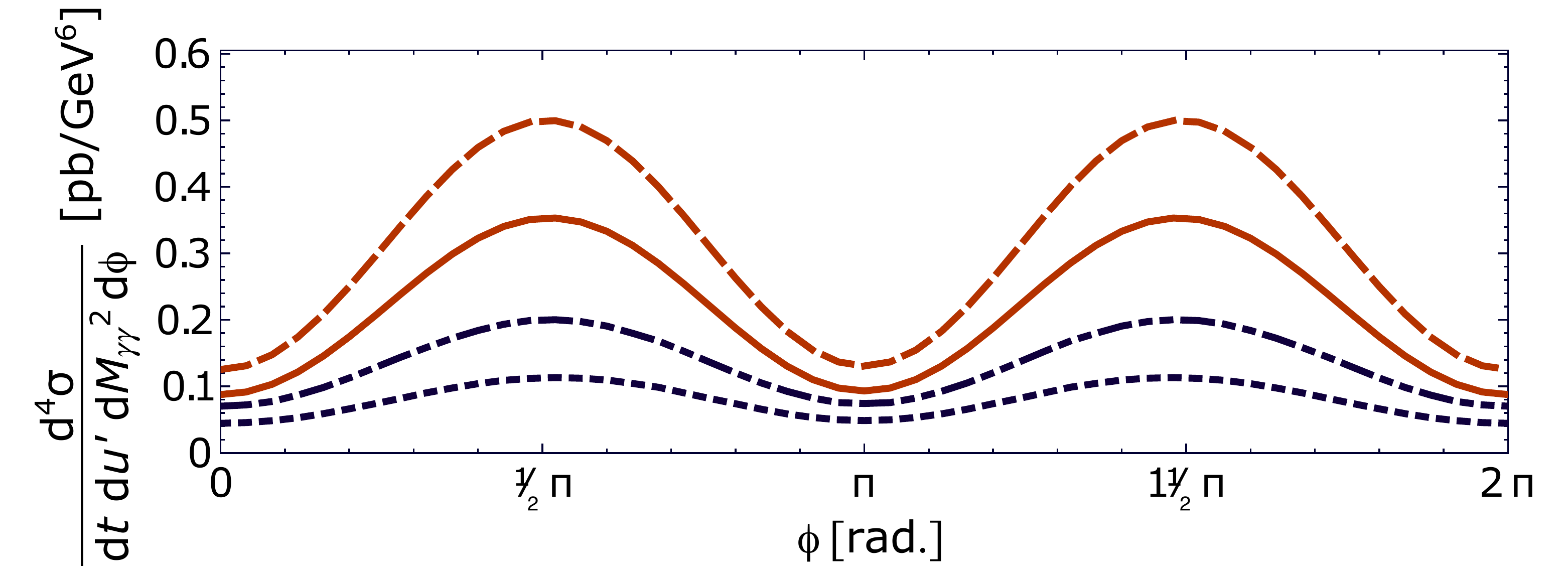}
\includegraphics[width=1.0\textwidth]{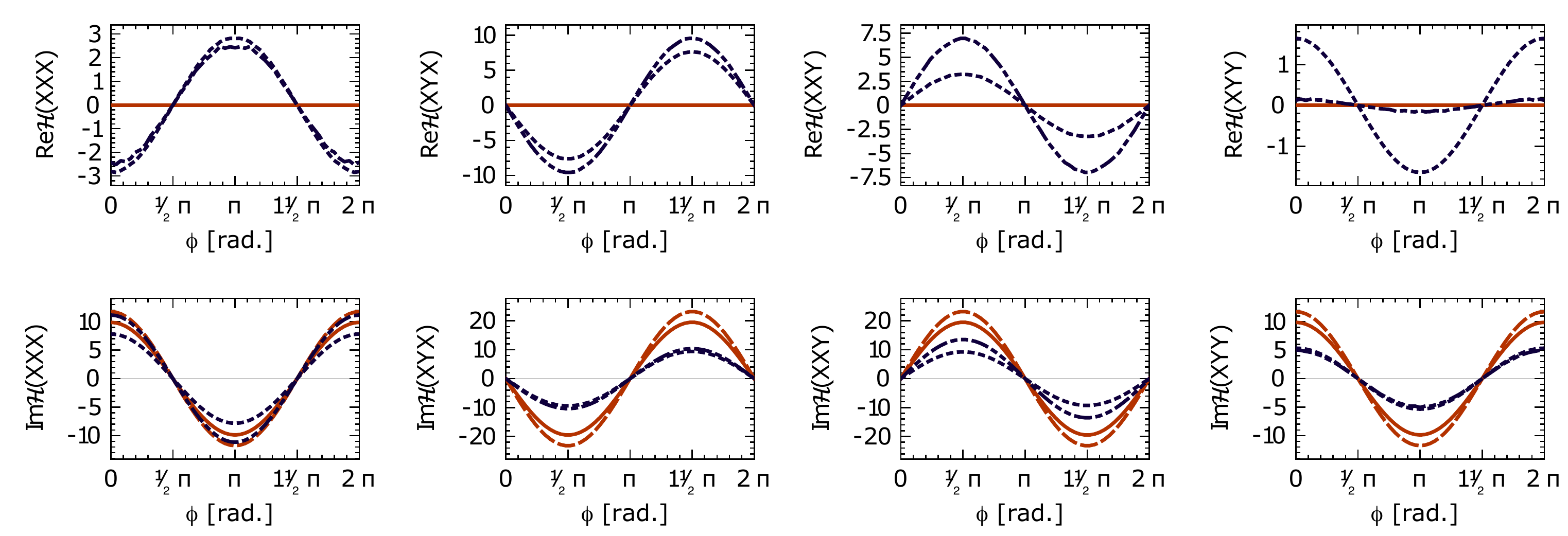}
\caption{(top) Differential cross-section as a function of $\phi$ -- the angle between the initial photon polarization and one of the final photon momentum in the transverse plane. (bottom) Compton form factors $\mathcal{H}$ as a function of $\phi$ for various combinations of polarisations. The plots are for $S_{\gamma N} = 20~\mathrm{GeV}^2$, $M_{\gamma\gamma}^2 = 4~\mathrm{GeV}^2$ (which corresponds to $\xi \approx 0.12$), $u' = -1~\mathrm{GeV}^2$ and $t=t_{0}\approx -0.05~\mathrm{GeV}^2$. For further description see the caption of Fig. \ref{fig:plot_1}.}
\label{fig:plot_4}
\end{center}
\end{figure}

As in the case of TCS \cite{Goritschnig:2014eba}, there is an interesting leading twist dependence on the initial photon polarization when this polarization is linear. We show on Fig. \ref{fig:plot_4}  the dependence of  the cross-section and CFFs on the angle $\phi$ between the transverse momentum $\vec{p}_\perp$ and transverse polarization component of the incoming photon for $S_{\gamma N} = 20~\mathrm{GeV}^2$, $M_{\gamma\gamma}^2 = 4~\mathrm{GeV}^2$ (corresponding to $\xi \approx 0.12$), $u' = -1~\mathrm{GeV}^2$ and $t=t_{0}\approx -0.05~\mathrm{GeV}^2$.

\subsection{Transversely polarized target asymmetry}
\label{sec:phenoT}

\begin{figure}[!ht]
\begin{center}
\includegraphics[width=0.5\textwidth]{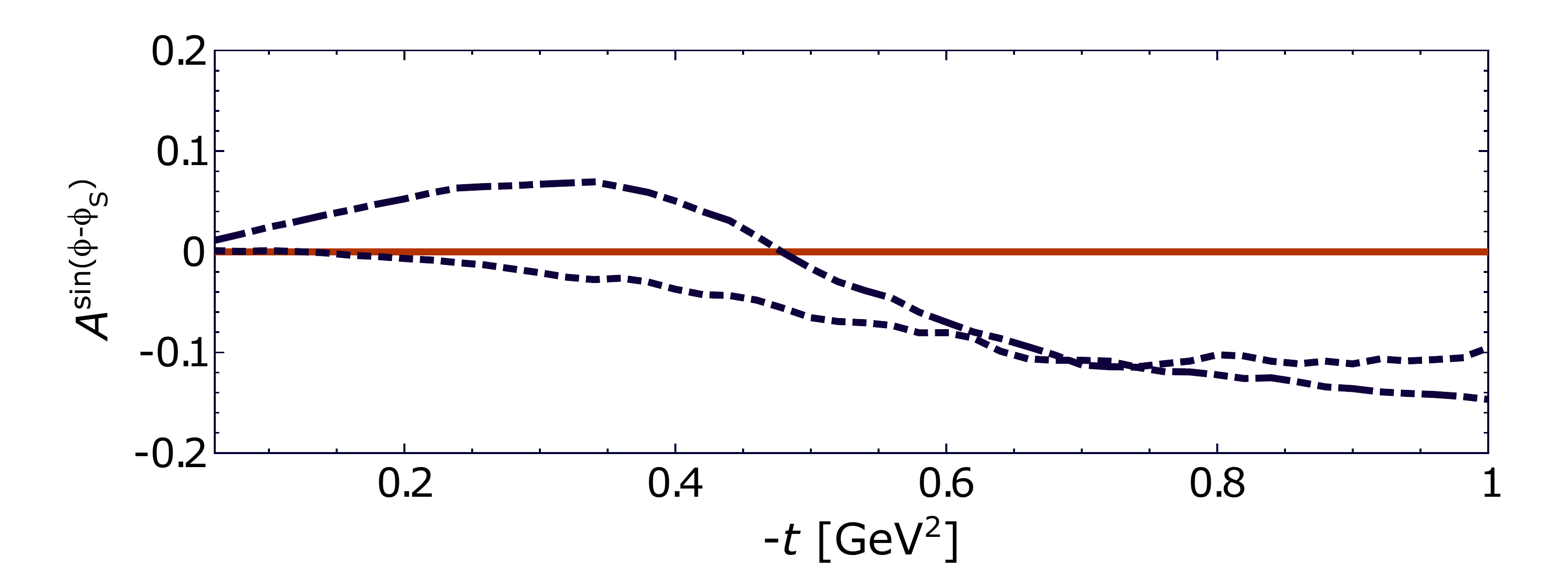}
\includegraphics[width=1.0\textwidth]{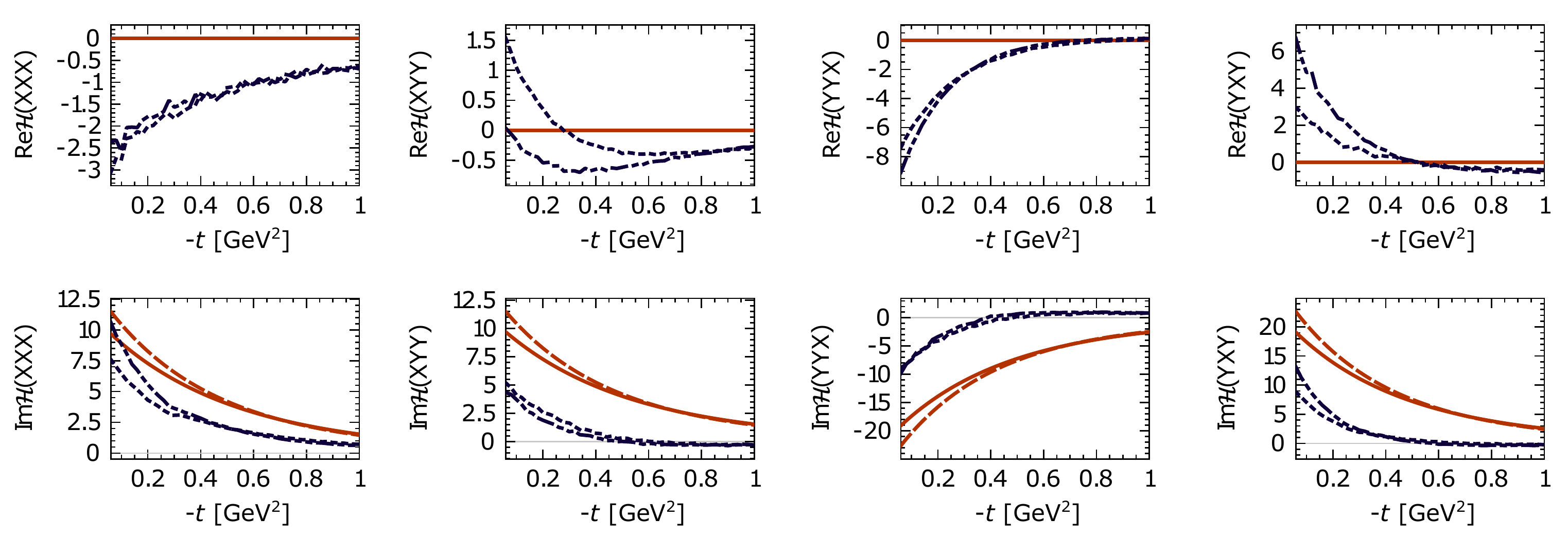}
\includegraphics[width=1.0\textwidth]{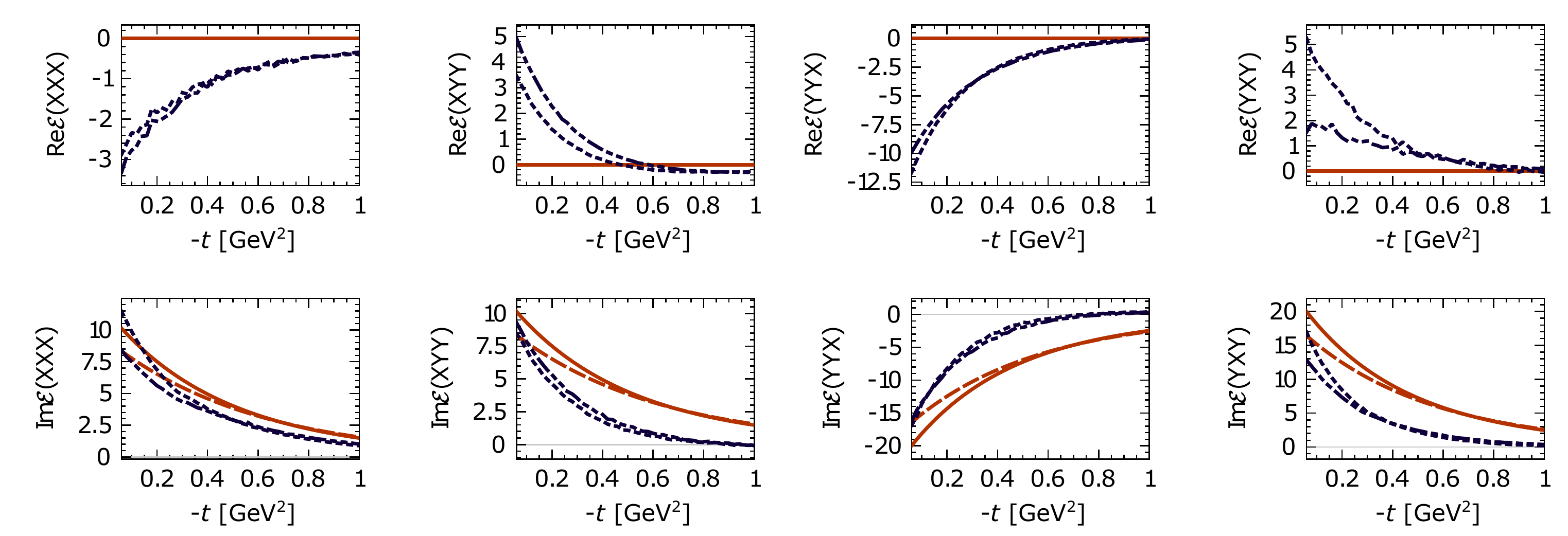}
\caption{(top) The transverse target asymmetry ${\cal A}^{\sin(\phi_{\Delta_T} - \phi_{S_T})}$ as a function of $-t$. (bottom) Compton form factors $\mathcal{H}$ and $\mathcal{E}$ as a function of $-t$ for various combinations of polarisations. The plots are for $S_{\gamma N} = 20~\mathrm{GeV}^2$, $M_{\gamma\gamma}^2 = 4~\mathrm{GeV}^2$ (which corresponds to $\xi \approx 0.12$) and $u' = -1~\mathrm{GeV}^2$. For further description see the caption of Fig. \ref{fig:plot_1}. Note that the asymmetry strictly vanishes in the LO case (red curves).}
\label{fig:plot_5}
\end{center}
\end{figure}

Similarly as in the case of  meson electroproduction, there is a leading twist single spin asymmetry when the nucleon target is transversely polarized. 
This target asymmetry  $\cal A$ is expressed in terms of CFFs $\cal H$ and $\cal E$ as:
\begin{eqnarray}
{\cal A} &=& \frac{\displaystyle\frac{\epsilon^{n\,p\,\Delta_T\,S_T}}{p\cdot n}\frac{(1+\xi)}{2M} \Im ({\cal E}^*{\cal H})}{\displaystyle{\cal H}{\cal H}^*(1-\xi^2) +\frac{\xi^4 }{1-\xi^2} {\cal E}{\cal E}^* +2\xi^2 \Re({\cal E}{\cal H}^*)} \nonumber
\\
&=& - \sin(\phi_{\Delta_T} - \phi_{S_T}) |\Delta_T| \frac{\displaystyle\frac{(1+\xi)}{2M} \Im ({\cal E}^*{\cal H})}{\displaystyle{\cal H}{\cal H}^*(1-\xi^2) +\frac{\xi^4 }{1-\xi^2} {\cal E}{\cal E}^* -2\xi^2 \Re({\cal E}{\cal H}^*)} \,.
\end{eqnarray}
Here, $\phi_{\Delta_T} - \phi_{S_T}$ is the relative angle between $\Delta_T$ and $S_T$, the latter being the transverse component of the target polarization vector, and:
\begin{equation}
\Delta_T^2 = \frac{1 - \xi}{1 + \xi} (t_0 - t) \,.
\end{equation}
The moment of this asymmetry:
\begin{equation}
{\cal A}^{\sin(\phi_{\Delta_T} - \phi_{S_T})} = \frac{1}{\pi} \int_0^{2\pi} \mathrm{d}(\phi_{\Delta_T} - \phi_{S_T}) {\cal A} \sin(\phi_{\Delta_T} - \phi_{S_T}) \,,
\end{equation}
is shown in Fig. \ref{fig:plot_5}, together with the corresponding distributions of both CFFs $\mathcal{H}$ and $\mathcal{E}$. The asymmetry vanishes at LO, since the real part of the CFFs vanishes at this order.
 The non-vanishing value of ${\cal A}^{\sin(\phi_{\Delta_T} - \phi_{S_T})}$ at NLO, even if predicted to not be large, thus directly probes both higher-order effects and the non-singlet contribution to GPD $E$, making this observable interesting for the extraction of this GPD. We note that the numerical noise is magnified in the asymmetry.

\section{Conclusions}
\label{sec:summary}

In this work we discussed various observables for the diphoton photoproduction
process: $\gamma N \rightarrow \gamma \gamma N $. The process is unique, as it probes $C$-odd combinations of GPD. Our estimates are made for either LO or NLO coefficients functions, and for two GPD models. Presented results suggest that experimental conditions of JLab experiments are optimal to measure this process. NLO corrections improve the accuracy of theoretical estimates for considered differential cross-sections and polarization asymmetries, and should be helpful in future planning of exclusive experiments on nuclei.

\clearpage

\paragraph*{Acknowledgements.}
\noindent
The work of O.G. is financed by the budget for science in 2020-2021, as a research project under the "Diamond Grant" program. The works of L.S. and J.W. are  supported respectively by the grants 2019/33/B/ST2/02588 and 2017/26/M/ST2/01074 of the National Science Center in Poland. This work is also partly supported by the Polish-French collaboration agreements Polonium, by the Polish National Agency for Academic Exchange and COPIN-IN2P3 and by the European Union’s Horizon 2020 research and innovation programme under grant agreement No 824093. The computing resources of {\'S}wierk Computing Centre, Poland are greatly acknowledged.

\bibliography{NLO_pheno}
\end{document}